\documentstyle[12pt]{article}
\def\Journal#1{{#1}}

\def\NIMA{{\em Nucl. Instrum. Methods} A}
\def\NPB{{\em Nucl. Phys.} B}
\def\NP{\em Nucl. Phys.}
\def\PLB{{\em Phys. Lett.}  B}
\def\PRL{\em Phys. Rev. Lett.}
\def\PRD{{\em Phys. Rev.} D}
\def\ZPC{{\em Z. Phys.} C}

\input psfig.sty

\begin{document}
\normalsize
\parskip=5pt plus 1pt minus 1pt
\title{\bf $R$-values in Low Energy $e^+e^-$ Annihilation}
\author{Zhengguo Zhao\\Institute of High Energy Physics of Chinese Acadmy of
Sciences\\Beijing 100039,P.R.C.}
%\date{  }
\maketitle

\begin{abstract} 
%The QED running coupling constant $\alpha(s)$ and the anomalous magnetic 
%moment of muon $a_{\mu}$ are two fundamental quantities for the precision 
%test of the Standard Model. The current uncertainties on $\alpha(s)$ and 
%$a_{\mu}$ are dominated by the contribution from the R-values measured 
%about 20 years ago with an averaged uncertainties of 15\% in the energy 
%region below 5 GeV.  
This presentation briefly summarizes the recent measurements of
$R$-values in low energy $e^+e^-$ annihilation. The new experiments
aimed at reducing the uncertainties in $R$-values and performed with the
upgraded Beijing Spectrometer (BESII) at Beijing Electron Positron
Collider (BEPC) in Beijing and with CMD-2 and SND at VEEP-2M in
Novosibirsk are reviewed and discussed.
\end{abstract}

\section{Introduction}

According to the quark-parton model, hadrons produced via $e^+e^-$
collisions are characterized by the annihilation of $e^+e^-$ pairs into
a virtual $\gamma$ or $Z^0$ boson. In the lowest order,
%the cross 
%section for the (QED) processes $e^+e^- \rightarrow q \bar q$ is 
%related to that for $e^+e^- \rightarrow \mu^+\mu^-$. 
%\begin{equation}
%\sigma(e^+e^- \rightarrow q \bar q)=3 Q^2_f \sigma(e^+e^- \rightarrow
%\mu^+\mu^-)
%\end{equation}
%where $Q_q$  is the fractional charge of the quark, and three in front 
%records the three colors for each flavor. Summing over all the quark flavors, 
one defines the ratio of the rate of hadron production to that for muon
pairs as
\begin{equation}
R \equiv \frac{\sigma (e^+e^- \rightarrow \mbox{hadrons})}{\sigma
(e^+e^- \rightarrow \mu^+\mu^-)}=3\sum_{f}Q^2_{f},
\end{equation}
where $Q_f$ is the fractional charge of the quark, and the factor of 
3in front
counts the three colors for each flavor.  $\sigma(e^+e^- \rightarrow
\mu^+\mu^-)=4\pi\alpha^2/3s$ is the cross section of the pure QED
process. The value of $R$, which counts directly the number of quarks,
their flavor and colors, is expected to be constant so long as the
center-of-mass (cm) energy of the annihilated $e^+e^-$ does not overlap
with resonances or thresholds for the production of new quark flavors.
One has
\begin{eqnarray}
R &=& 3[(2/3)^2+(1/3)^2+(2/3)^2] = 2~~     \mbox{for}~~u, d, s \nonumber \\ 
  &=& 2      + 3(2/3)^2        = 10/3~~    \mbox{for}~~u, d, s, c, \nonumber \\
  &=& 10/3 + 3(1/3)^2          = 11/3~~    \mbox{for}~~u, d, s, c, b. \nonumber
\end{eqnarray}

These values of $R$ are only based on the leading order process $e^+e^-
\rightarrow q \bar q$. However, one should also include the
contributions from diagrams where the quark and anti-quark radiate
gluons.  The higher order QCD corrections to $R$ have been calculated in
complete 3rd order perturbation theory \cite{Rhighorder}, and the
results can be expressed as
\begin{equation}
R=3\sum_{f}Q^2_{f}[1+(\frac{\alpha_s(s)}{\pi})+1.411(\frac{\alpha_s(s)}{\pi})^2
-12.8(\frac{\alpha_s(s)}{\pi})^3+\ldots],
\end{equation}
where $\alpha_s(s)$ is the strong coupling constant. Precise measurement
of $R$ at higher energy can be employed to determine $\alpha_s(s)$
according to eq.2, which exhibits a QCD correction known to
$O(\alpha^3_s)$ . In addition, non-perturbative corrections could be
important at low cm energy, particularly in the resonance region.

\begin{figure}[htbp]
\centerline{
\psfig{figure=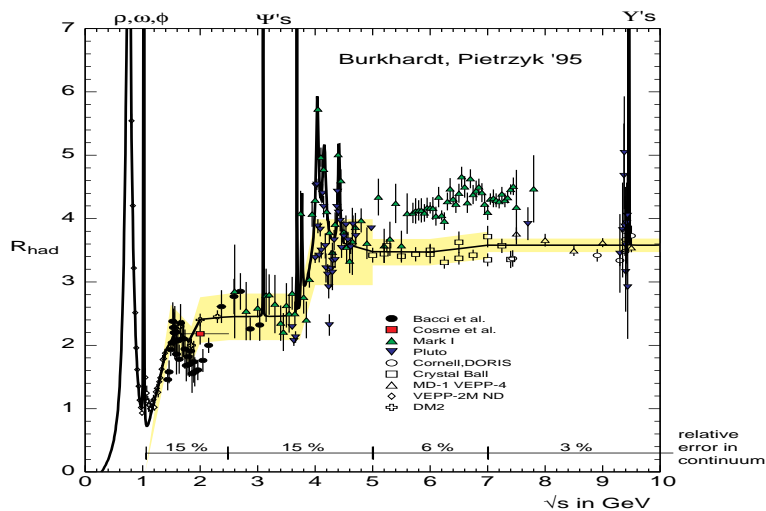,width=100mm,height=66mm}}
\caption{Experimental $R$ values in the energy region below 10 GeV. The
relative errors on $R$ in the continuum is given in numbers at 
the bottom of the figure.}
\label{fig:Rto10}
\end{figure}

$R$ has been measured by many laboratories in the energy region from
hadron production threshold to the $Z^0$ pole and recently to the energy
of W pair production \cite{RtoZ}. The experimental $R$-values are in
general consistent with theoretical predictions, which is an impressive
confirmation of the hypothesis of the three color degrees of freedom for
quarks.  The measurements of $R$ in the low energy region were performed
15 to 20 years ago in Novosibirsk, Orsay, Frascati, SLAC and Hamburg
~\cite{Rlow1,Rlow2,Rlow3,Rlow4,gamma2,MarkI,pluto}.
Fig.~\ref{fig:Rto10}~\cite{BP} shows the $R$-values for cm energies up
to 10 GeV, including resonances.  For cm energies below 5 GeV, the
uncertainties in $R$-values are about ~15\% on average; and the
structure in the charm threshold region is not well determined.  The DASP
group~\cite{DASP} inferred the existence of narrow resonances at 4.04 GeV
and 4.16 GeV. In addition to the resonance at 3.77 GeV, Mark I
data~\cite{MarkIres} shows broad enhancements at 4.04, 4.2 and 4.4
GeV. The resonance at 4.4 GeV was also observed by
PLUTO~\cite{PLUTOres}, but the height and width of the resonance were
reported differently.  A new cross section measurement in the charm
threshold region is needed to clarify the structure, which is
important not only for the precision determination of $\alpha(M^2_Z)$
and the interpretation of the $(g-2)_{\mu}$ measurement of E821 
at BNL but also for the understanding of charmonium itself.

Between the charm and bottom thresholds, i.e., about 5-10.4 GeV, $R$ was
measured by Mark I, DASP, PLUTO, Crystal Ball, LENA, CLEO, CUSB, and
DESY-Heidelberg collaborations. Their systematic normalization
uncertainties were about 5-10\%. Above bottom threshold, the
measurements were from PEP, PETRA and LEP with uncertainties of 2-7\%.
\begin{table}[h]
\caption {R in low energy measured by different laboratories}
\vskip 0.5cm
\begin{center}
\begin {tabular}{|c|c|c|c|c|c|} \hline
Place      & Ring   & Detector            & $E_{cm}$  & Points & Year \\
           &        &                     &  (GeV)    &        &      \\
\hline
Beijing    & BEPC   & BESII               &  2.0-5.0  &   85   & 1998-1999 \\
\hline
Novosibirsk&VEPP-2M & CMD-2               & 0.6-1.4   &  128   & 1997-1999 \\
           &        & SND                 &           &        &           \\
           & VEPP-2 & Olya,ND CMD         & 0.3-1.4   &        &           \\
\hline
SLAC       & Spear  & MarkI               & 2.8-7.8   &   78   & 1982      \\
\hline
Frascati   & Adone  & $\gamma\gamma2$,MEA & 1.42-3.09 &   31   & 1978      \\  
           &        & Boson,BCF           &           &        &           \\
\hline
Orsay      & DCI    & M3N,DM1,DM2         & 1.35-2.13 &   33   & 1978      \\
\hline
Hamburg    & Doris  & DASP                & 3.1-5.2   &   64   & 1979      \\ 
           &        & PLUTO               & 3.6-4.8   &   27   & 1977      \\
\hline
\end{tabular}
\end{center}
\label{tab:R_value}
\end{table}

Remarkable progress has been made in the precision test of the Standard
Model (SM) during the last decade. The electroweak data from LEP is so
copious and precise that one can make use of the radiative correction
effect to test the SM. In particular, the indirect determination of
$m_H$ depends critically on the precision of $\alpha(M^2_Z)$. Recently,
for example, there has been an increasing interest in electroweak
phenomenology to reduce the uncertainty in $\alpha(M_Z^2)$ which
seriously limits further progress in the determination of the Higgs mass
from radiative corrections to the
SM~\cite{BP,Bolek,Davier1,Eidelman,swartz,Davier2}.
%For the analysis of electroweak data in the
%SM~\cite{Glashow,Weinberg} one starts from the input parameters. Some of 
%them, like $\alpha(M^2_Z)$, $G_F$, and $M_Z$ are very well known, and some 
%others, $m_{light}$, $\alpha_s(M^2_Z)$ and $m_t$ are only approximately 
%determined, whereas $m_H$ is almost unknown. Constrain on $m_t$ and $m_H$
%can be derived by comparing the measured observables with theoretical 
%predictions that has been calculated to full one-loop accuracy and partial
%two-loop precision, a sufficient precision to match the experimental 
%capabilities.
%Out of the three accurately determined quantities $\alpha(s)$, $G_F$, and
%$M_Z$, the largest uncertainty comes from the running of QED coupling
%constant $\alpha(s)$ from s = 0, where it is known to 0.04 ppm, up to the
%Z pole, which is the scale relevant for the electroweak precision test.
%The running of $\alpha(s)$ as a function of $s= \sqrt{q^2}$ is shown in
%Fig. 2. 
%When relating measurements performed at different energy scales,
%and if the relation involves $\alpha(s)$, one has to know the running of
%$\alpha(s)$ in different energy scale.
The uncertainty in $\alpha(M^2_Z)$ arises from the contribution of light
quarks to the photon vacuum polarization $\Delta \alpha(s)=
-\prod_{\gamma}^{'}(s)$ at the Z mass scale.  They are independent of
any particular initial or final states and can be absorbed in $\alpha(s)
\equiv \alpha/[1-\Delta \alpha(s)]$, where the fine-structure constant
$\alpha$=1/137.035 989 5(61). $\Delta \alpha = \Delta \alpha_{lepton} +
\Delta \alpha_{had}$, of which the leptonic part is precisely
calculated analytically according to perturbation theory because free
lepton loops are affected by small electromagnetic
corrections~\cite{Jegerlehner}.  Whereas, the hadronic part $\Delta
\alpha_{had}$ cannot be entirely calculated from QCD because of
ambiguities in defining the light quark masses $m_u$ and $m_d$ as well
as the inherent non-perturbative nature of the problem at small energy
scale since the free quark loops are strongly modified by strong
interactions at low energy.  An ingenious way to handle this[10] is 
to relate $\Delta \alpha_{had}$ from the quark loop diagram to $R$, 
making use of unitarity and analyticity,
\begin{equation}
\Delta\alpha_{had}^{(5)}(s) = -\frac{s}{4\pi^2\alpha} P(     
\int_{4m_{\pi}^2}^{E^2_{cm}}ds'\frac{R^{data}(s')}{s'(s'-s)}
+\int_{E^2_{cm}}^{\infty}ds'\frac{R^{PQCD}(s')}{s'(s'-s)}),
\end{equation}
where $R(s)=\sigma(e^+e^- \rightarrow hadrons)/\sigma(e^+e^- \rightarrow
\mu^+\mu^-)$
%= 12 \pi lm \Pi_{\gamma}^{'}(s)
%\begin{equation}
%R(s)=\sigma(e^+e^- \rightarrow hadrons)/\sigma(e^+e^- \rightarrow \mu^+\mu^-) 
%= 12 \pi lm \Pi_{\gamma}^{'}(s)
%\end{equation}
and P is the principal value of the integral.

Much independent work has recently been done to evaluate
$\alpha(s)$ at the energy of the Z pole. So far, the uncertainty of $\Delta
\alpha(s)$ is dominanted by the $R$-values at low energy ($E_{cm}<5~GeV$)
measured with an average uncertainty of $\sim$15\%, as indicated in
Fig.~\ref{fig:Rto10}.
%Recently there is a theoretical driven tendency to 
%extend the PQCD to the energy down to 1.7 GeV in order to avoid using the 
%experimental R-values measured 20 years ago with large uncertainties. 
%However, such a tendency
%is not deter the experimentalists to improve R measurements, which can
%test such a kind of calculation base on QCD. 
%Table ~\ref{tab:table2}. 
%summarized the evaluated $\alpha(M^2_Z)$ values by different authors.

The anomalous magnetic moment of the muon $a_{\mu}\equiv(g-2)/2$
receives radiative contributions that can in principle be sensitive to
new degrees of freedom and interactions. Theoretically, $a_{\mu}$ is
sensitive to large energy scales and very high order radiative
corrections.
%~\cite{g2rad1,g2rad2}. 
It therefore provides an extremely clean test of electroweak theory and
may give us hints on possible deviations from the SM.
%~\cite{g2SM1,g2SM2,g2SM3}. 
The experimental and the theoretical predictions on $a_{\mu}$ are well
reviewed by Dr. Robert Lee in his talk given at LP99.
%The weak interaction and the vacuum polarization
%effects are too small to observe for electron because of $m_l^2$-dependence.
%The measurement of $a_{\tau}$ is very difficult due to its short lifetime.
%However, benefited from its larger mass and relatively long lifetime the
%anomalous magnetic moment of muon 
%$a_{\mu}$ has been measured with very high
%precision at the CERN Muon Storage Ring~\cite{g21,g22,g23}, which 
%is one of the best measured quantities in physics. Theoretically, 
%$a_{\mu}$ is sensitive to large energy scales and very high order radiative
%corrections~\cite{g2rad1,g2rad2}.
%It therefore provides an extremely clean test of electroweak theory and may
%give us hints on possible deviations from the SM~\cite{g2SM1,g2SM2,g2SM3}. The
%experimental and the theoretical prediction on $a_{\mu}$ are well reviewed 
%by Dr. Robet Lee in his talk given at LP99.

One can decompose $a_{\mu}$ as
\begin{equation}
a_{\mu}^{SM} = a_{\mu}^{QED}+a_{\mu}^{had}+a_{\mu}^{weak},
\end{equation}
where the largest term, the QED contribution $a_{\mu}^{QED}$, has been
calculated to $O(\alpha^5)$, including the contribution from $\tau$
vacuum polarization.  $a_{\mu}^{weak}$ includes the SM effects due to
virtual W, Z and Higgs particle exchanges. $a_{\mu}^{had}$ denotes the
virtual hadronic (quark) contribution determined by QCD, part of which
corresponds to the effects representing the contribution of the running
of $\alpha(s)$ from low energy to high energy scale.  It cannot be
calculated from first principles but can be related to the experimentally
determined $R(s)$ through the expression
\begin{equation}
a^{had}_{\mu}=(\frac{\alpha m_{\mu}}{3\pi})^2
\int_{4m_{\pi}^2}^{\infty}ds\frac{R(s)K(s)}{s^{2}},
\end{equation}
where $K(s)$ is a kernel varying from 0.63 at $s=4 m^2_{\pi}$ to 1.0 at
$s= \infty$.

The hadronic vacuum polarization is the most uncertain of all the SM
contributions to $a_{\mu}$, the uncertainty is presently 
$156 \times 10^{-11}$. For several
scenarios, it has been claimed that ``the physics achievement of the
effort to measure the cross section of $e^+e^- \rightarrow
\mbox{hadrons}$ that brings down the uncertainty of $a_{\mu}$ to $60
\times 10^{-11}$ is equivalent to that of LEP2 or even the 
LHC''~\cite{Blondel}.

From equations 3 and 5 one finds that $a^{had}_{\mu}$ is more sensitive
to lower energies than higher ones.  Further measurement in the energy
region of 0.5-1.5 GeV from VEPP-2M in Novosibirsk and DA$\Phi$NE in
Frascati will contribute to the interpretation of the $a_{\mu}$ measurement
at Brookhaven~\cite{Lee} and the luminosity measurement at
CERN~\cite{Bolek}. However, their contribution to the precision
determination of $\alpha(M_Z^2)$ is limited. The improved $R$ value from
BESII at BEPC in the energy region of 2-5 GeV will make the major
contribution to evaluate $\alpha(M_Z^2)$, and also partly contribute to
the interpretation of $a_{\mu}$.
%Fig.~\ref{fig:un_amu_alpha} shows the relative contributions to the
%uncertainties of $a_{\mu}$ and $\alpha(M^2_Z)$.
%\begin{figure}[htb]
%\centerline{
%\psfig{figure=eps/un_amu_alpha.eps,height=6cm}}
%\caption{The relative contribution of the uncertainties in (a) $a_{\mu}$ and 
%(b)$\Delta \alpha(M^2_Z)$ ~\cite{Davier1}}
%\label{fig:un_amu_alpha}
%\end{figure}

\section{Recent measurements of $R$ in low energy $e^+e^-$}

There are two different approaches to the measurement of $R$. One is to
study the exclusive hadronic final states, i.e. to measure the
production cross section of each individual channel $\sigma^{exp}(e^+e^-
\rightarrow \mbox{hadrons})_j$. The value of $R$ can then be obtained by
summing over the measured hadron production cross section of all
individual channels.  This method demands that the detector has good
particle identification and requires the understanding of each channel.
It is usually used for cm energies below 2 GeV.
%Fig.~\ref{fig:hadthr} indicates different individual channels and their
%production thresholds.
%\begin{figure}[htb]
%\vspace{0.5cm}
%\centerline{
%\psfig{figure=eps/hadthr.eps,height=60mm,clip=true}}
%\caption{Production threshold of different hadronic final states}
%\label{fig:hadthr}
%\end{figure}

Another method treats the hadronic final states inclusively. It measures
$R$ by dealing with all the hadronic events simultaneously and is suitable
in an energy region where a reliable event generator for hadron
production is available. With an improved Lund Model~\cite{bo}, we may
be able to extend this region down to 2 GeV.
%This method relies on 
%MC generator to obtain acceptance-corrected values of R. Experimentally, the 
%R value is defined as 
%\begin{equation}
%R=\frac{\sigma^0_{had}(s)}{\sigma^0_{\mu\mu}(s)} 
%=\frac{1}{\sigma^0_{\mu\mu}} \cdot \frac{N_{had}-N_{bg}}{L \cdot 
%\varepsilon_{had} \cdot (1+\delta)}
%\end{equation}
%where $N_{had}$ and $N_{bg}$ are the observed hadronic events and all kind
%of background respectively; L represents the integrated luminosity of the
%colliding beam; $\delta$ is the radiative correction to hadron production,
%and $\varepsilon_{had}$ the detection efficiency for the hadronic events.
%$\sigma_{\mu \mu}^0= 4\pi \alpha^2/3s$

The typical features of hadron production below 5 GeV are:\\
\noindent $\bullet$ many resonances in this energy region, such as,
$\rho$, $\omega$, $\varphi$,  $\rho'$, $\omega'$, $\varphi'$, 
$c \bar c$ and charmed mesons $J/\psi$, $\psi(2S)$, $D^+D^-$,
$D_s^+D_s^-$ and $\tau^+\tau^-$, baryon-antibaryon pair production \\
$\bullet$ a small number of final states and low charged multiplicity,
usually $N_{ch} \le 6$.\\
The experimental challenge here is how to subtract the beam-associated
background and select $N_{had}$.

In the following section, I will first discuss some new measurements
done by CMD-2 and SND at VEPP-2M in Novosibirsk, which are based on an
exclusive analysis of the hadron production in the energy region around
0.4-1.4 GeV. Then I will concentrate on discussing the $R$ scan done by
BESII at BEPC in Beijing in the energy region from 2-5 GeV, which
measures $R$ values by dealing with hadronic final states inclusively.

\subsection{Recent results from VEPP-2M}

VEPP-2M, the $e^+e^-$ collider with maximum luminosity of $\sim 5 \times
10^{30}$ cm$^{-2}$s$^{-1}$ at $E_{beam}$=510 MeV, has been operating
since 1974 in the energy region $E_{cm}=0.4-1.4$ GeV ($\rho$, $\omega$,
$\phi$-meson region). SND~\cite{snddet} and CMD-2~\cite{cmd2det} are the
two detectors carrying out experiments at VEPP-2M.  Since 1994, VEPP-2M
performed a series of scans from 0.38 GeV to 1.38 GeV
~\cite{vepp2mscan}. With this data both SND and CMD-2 have measured the
cross sections for the channels $ \pi^+ \pi^-$, $\pi^+ \pi^- \pi^0$,
$\pi^+\pi^-\pi^+\pi^-$, $\pi^+\pi^-\pi^0\pi^0$,
$\pi^+\pi^-\pi^+\pi^-\pi^0$ as well as $K_L K_S$ and $K^+ K^-$.

%\subsubsection{Recent results from SND}
%SND is described in detail in ref.\cite{snddet} 
%and ploted in Fig.~\ref{fig:snddet}. 
%The three layer spherical 
%electromagnetic calorimeter consisting of 1620 $NaI(TI)$ crystals with 
%total mass of 3.6 tones is its main part. The calorimeter covers 90\% of 
%$4\pi$ steradian solid angle. Its energy resolution for the photons is  
%$\sigma_E(E)/E = 4.2\%/E (GeV)^{1/4}$, and the angular resolution is 
%$\sim 1.5^0$. The two cylindrical drift chambers covering 95\% of $4\pi$ solid 
%angle measures the angle of the charged particles to the accuracy of  
%$\sim 0.4^0$ and $ \sim2.0^0$ in azimuth and polar direction respectively. As 
%an low energy neutral detector, SND is good at detecting $\gamma$. Following
%are some recent results from SND~\cite{sndresult}. 	
%
%\begin{figure}[htb]
%\centerline{
%\psfig{figure=eps/snddet.eps,height=60mm,clip=true}}
%\caption{SND detector}
%\label{fig:snddet}
%\end{figure}

\noindent $\bullet$ Study of $e^+e^-\rightarrow \pi^+\pi^-\pi^+\pi^-,
\pi^+\pi^-\pi^0\pi^0$

The four pion final states produced via $e^+e^-$ annihilation in the
$E_{cm}=1 \sim 2$~GeV energy region dominate and determine the main
part of the hadronic contribution to $a_{\mu}$ and the QCD sum
rules. Besides, these processes are important sources of information for
the understanding of hadron spectroscopy, in particular for the study of
the $\rho$-meson radial excitation~\cite{hadronspe}.  These processes
were studied at the VEPP-2M, DCI and ADONE colliders
~\cite{snd15,snd47,snd48,snd49,snd50}. The statistical errors of these
measurements were $\sim5\%$, and the systematic errors were $\sim15\%$.  
There was about a 20\% discrepancy among the different experiments.

\begin{figure}[htbp]
  \begin{minipage}[t]{0.48\linewidth}
    \centerline{\psfig{file=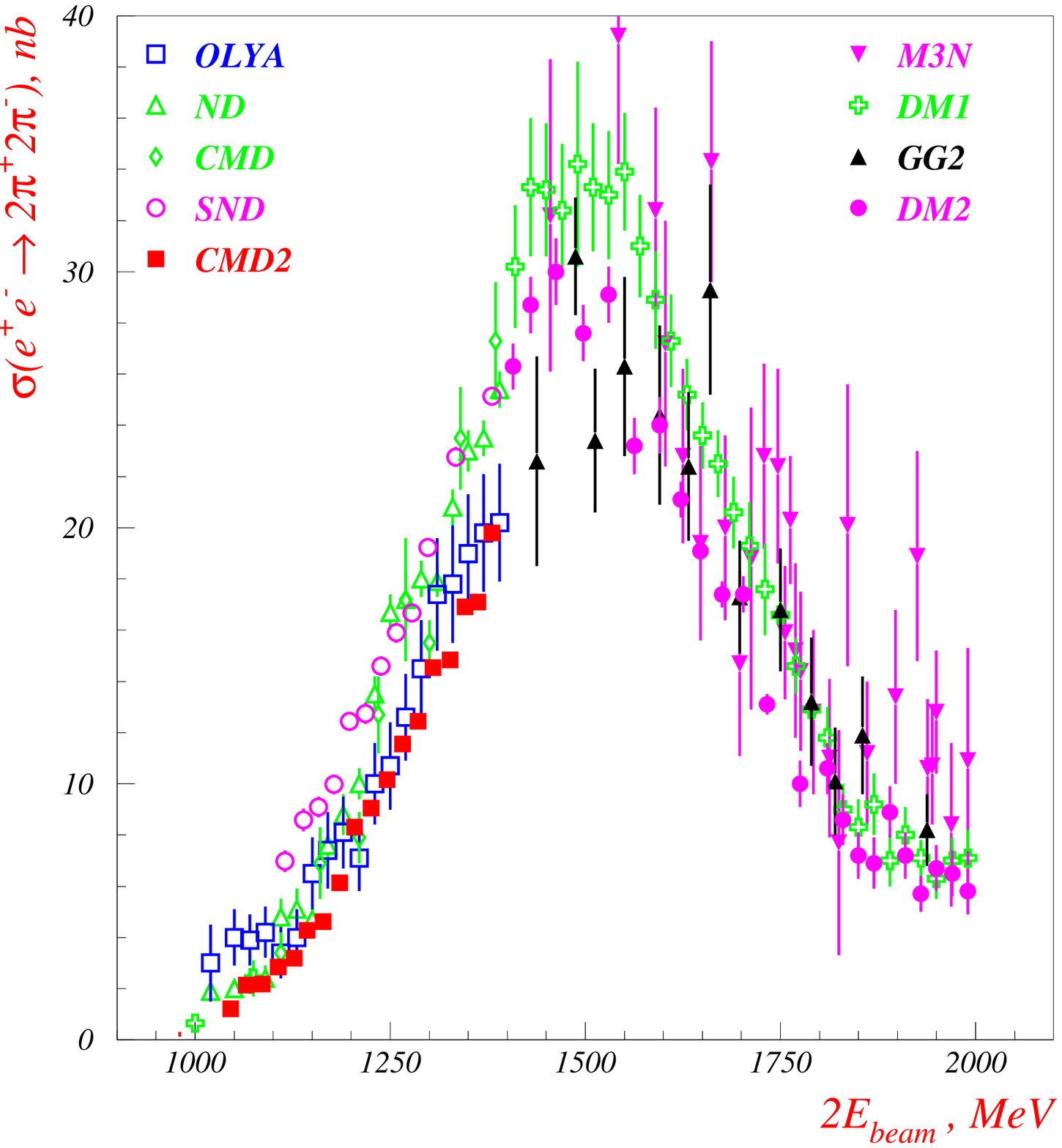,width=2.2in}}
    \caption[]{Energy dependence of the cross section of
               $e^+e^-\rightarrow2\pi^+ 2\pi^-$.}
    \label{fig:4pi}
  \end{minipage}
  \hfill
  \begin{minipage}[t]{0.48\linewidth}
    \centerline{\psfig{file=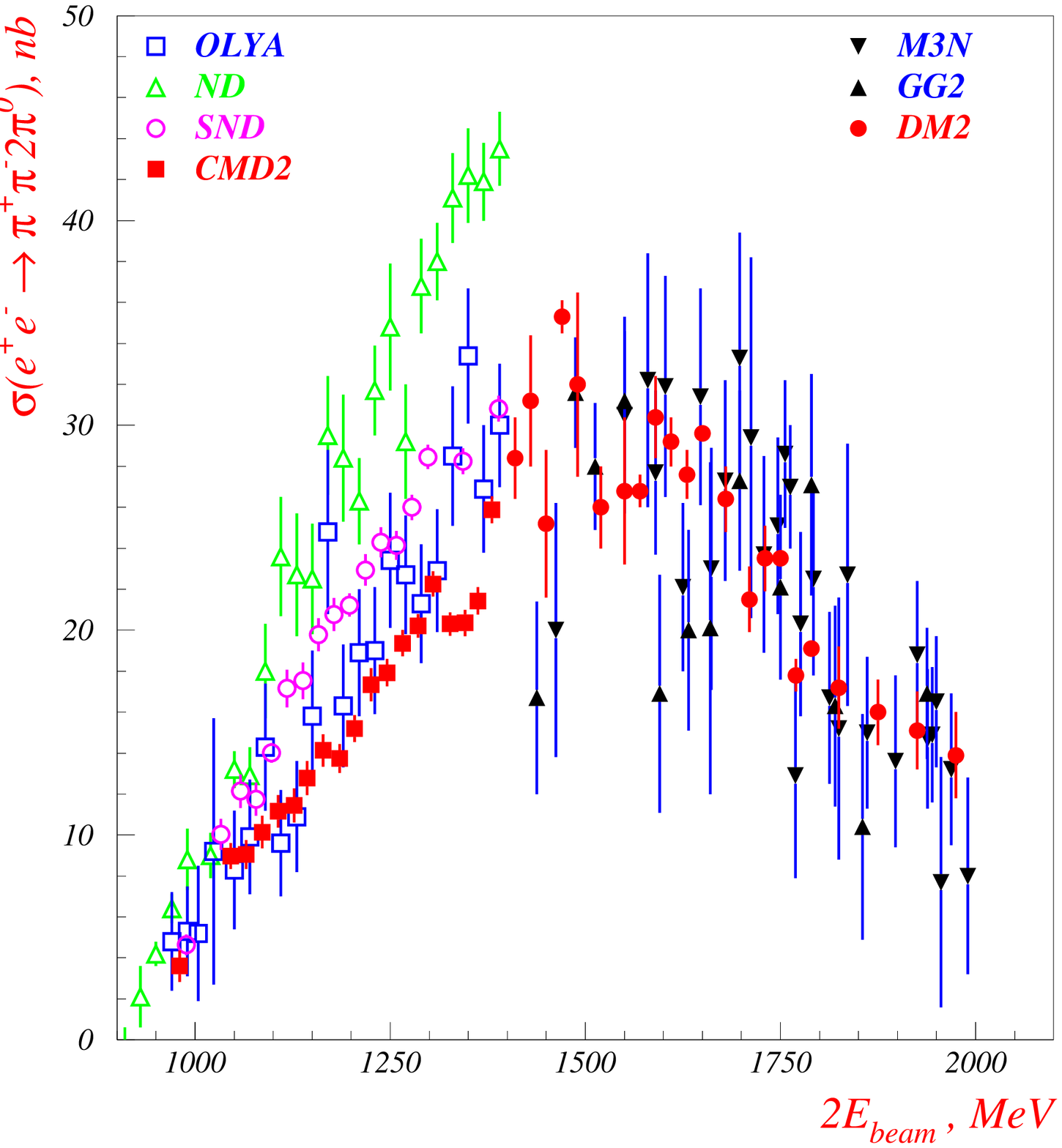,width=2.2in}}
    \caption[]{Energy dependence of the cross section of
               $e^+e^-\rightarrow\pi^+\pi^-2\pi^0$.}
    \label{fig:2pi2pin}
  \end{minipage}
\end{figure}

For the process of $e^+e^-\rightarrow \pi^+\pi^-\pi^+\pi^-$, the 
systematic error from the recent SND results is $\sim7\%$, mainly 
coming from the event selection and the luminosity determination. 
The measured total cross sections from CMD-2 for $e^+e^-\rightarrow 
2\pi^+ 2\pi^-$ are also illustrated in Fig.~\ref{fig:4pi}. Only the 
statistical errors are shown. The systematic uncertainties are 
$\sim7\%$, attributed to the luminosity measurement, the event 
reconstruction and selection and the radiative correction as well. 

For SND, backgrounds to the $e^+e^-\rightarrow \pi^+\pi^-\pi^0\pi^0$
channel are mainly from $e^+e^- \rightarrow K^+K^-$, QED processes
$e^+e^- \rightarrow e^+e^- e^+e^-, e^+e^-\gamma\gamma$, as well as
cosmic rays and beam associated background. The systematic error is
$\sim7\%$, of which $\sim5\%$ arise out of the variation of the
detection efficiency shown from the simulation of the intermediate
states like $\omega \pi^0$, $\rho^0 \pi^0 \pi^0$ and Lorentz-invariant
phase space simulation(LIPS).

The total cross section for $e^+e^-\rightarrow \pi^+ \pi^- 2\pi^0$
process measured by CMD-2 also is plotted in Fig.~\ref{fig:2pi2pin}.
The error bars indicates only the statistical error. The systematic
uncertainties mainly come from event reconstruction, radiative
corrections and the luminosity determination. The overall systematic
uncertainty is estimated to be 7\%. The cross section measured by this
experiment is consistent with what was measured by OLYA~\cite{olya} 
and a recent result from SND~\cite{Benayoun}.  However, the cross section
from all three measurements is apparently lower than that given by
ND~\cite{cmd223,cmd224}.  For comparison, the results from Orsay and
Frascati above 1.4 GeV are also shown in the figure.

CMD-2 finds that the dominant contribution to the cross section of the
process $e^+e^-\rightarrow \pi^+ \pi^- 2\pi^0$ comes from $\omega \pi^0$
and $\rho^{\pm} \pi^{\mp} \pi^0$ intermediate states, whereas the
$\rho^{\pm} \pi^0 \pi^0$ state is not observed. The $\rho^{\pm}
\pi^{\mp} \pi^0$ states are saturated completely by the $a_1(1260) \pi$
intermediate state. This is also the dominant contribution to the cross
section for the process $e^+e^- \rightarrow 2\pi^+ 2\pi^-$. The
theoretical predictions for the differential distributions and the total
cross sections can be dramatically changed if one takes into account the
interference of different amplitudes with various intermediate states
but identical final states.

The cross section of $e^+e^-\rightarrow 4\pi$ can be related to the
the four $\pi$ decays of the
$\tau$-lepton through the hypothesis of CVC~\cite{Eidelman}. This has
been experimentally tested to be valid within an accuracy of
3-5\%~\cite{Brown}. The observed $a_1(1260) \pi$ dominance, if it is
true, should be taken into account in $\tau$ decays.
%Fig.~\ref{fig:4pi} and Fig.~\ref{fig:2pi2pin} show 
%the cross sections measured from SND and CMD-2 for $2\pi^+2\pi^-$ and 
%$\pi^+\pi^-\pi^0\pi^0$.
%    \centerline{\psfig{file=eps/4pi_1-ok.eps,width=2.2in}}
%    \centerline{\psfig{file=eps/2pi2pin-ok.eps,width=2.2in}}

%\begin{figure}[htb]
%%\begin{center}
%\mbox{
%\psfig{figure=eps/4pi_1.eps,width=70mm,angle=90,clip=true}}
%\mbox{
%\psfig{figure=eps/2pi2pin.eps,width=70mm,angle=90,clip=true}}
%\parbox[l]{6.cm}{
%\caption{Energy dependence of the cross section of $e^+e^- \rightarrow 
%2\pi^+ 2\pi^-$.}
%\label{fig:4pi}}
%\hspace{1.6cm}
%\parbox{6.cm}{
%\caption{Energy dependence of the cross section of $e^+e^- \rightarrow 
%\pi^+\pi^-2\pi^0$.}
%\label{fig:2pi2pin}}
%\end{center}
%\end{figure}

\noindent $\bullet$  The investigation of $e^+e^-\rightarrow\pi^+\pi^-\pi^0$
	
This process was measured by ND at VEPP-2M in the energy region up to
1.1 GeV~\cite{dolinsky}. The measured cross section is significantly
higher than that predicted by the Vector Dominance Model (VDM). However,
it is well known that the VDM is able to well describe the cross section
near the $\omega$ and $\phi$ resonances for the processes
$e^+e^-\rightarrow\omega,~\phi
\rightarrow\rho\pi\rightarrow\pi^+\pi^-\pi^0$. It is therefore necessary
to perform new precise measurements in the non-resonance region to
investigate the limitation of the VDM and determine possible contributions
from heavier intermediate states like $\omega(1120)$ or $\omega(1600)$.
	 
SND also measured this channel. The systematic errors from the detection
efficiency, luminosity measurement and the background subtraction are
10\%, 5\% and 5\% respectively, giving a total systematic error of
$\sim12\%$.  The results from the new measurement agrees with the old ND
data.

\begin{figure}[htbp]
  \begin{minipage}[t]{0.48\linewidth}
    \centerline{\psfig{file=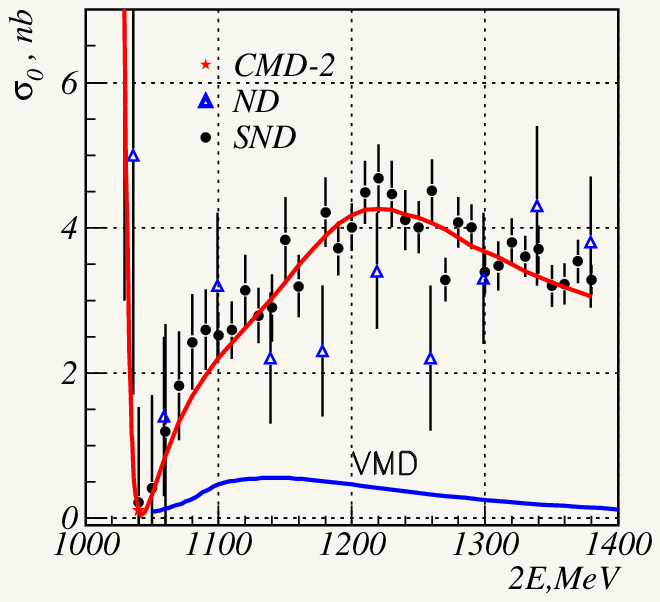,width=60mm}}
    \caption[]{Energy dependence of the cross section of 
               $e^+e^-\rightarrow \pi^+\pi^-\pi^0$.
               Solid line shows the VDM prediction.}
    \label{fig:pipipi0}
  \end{minipage}
  \hfill
  \begin{minipage}[t]{0.48\linewidth}
    \centerline{\psfig{file=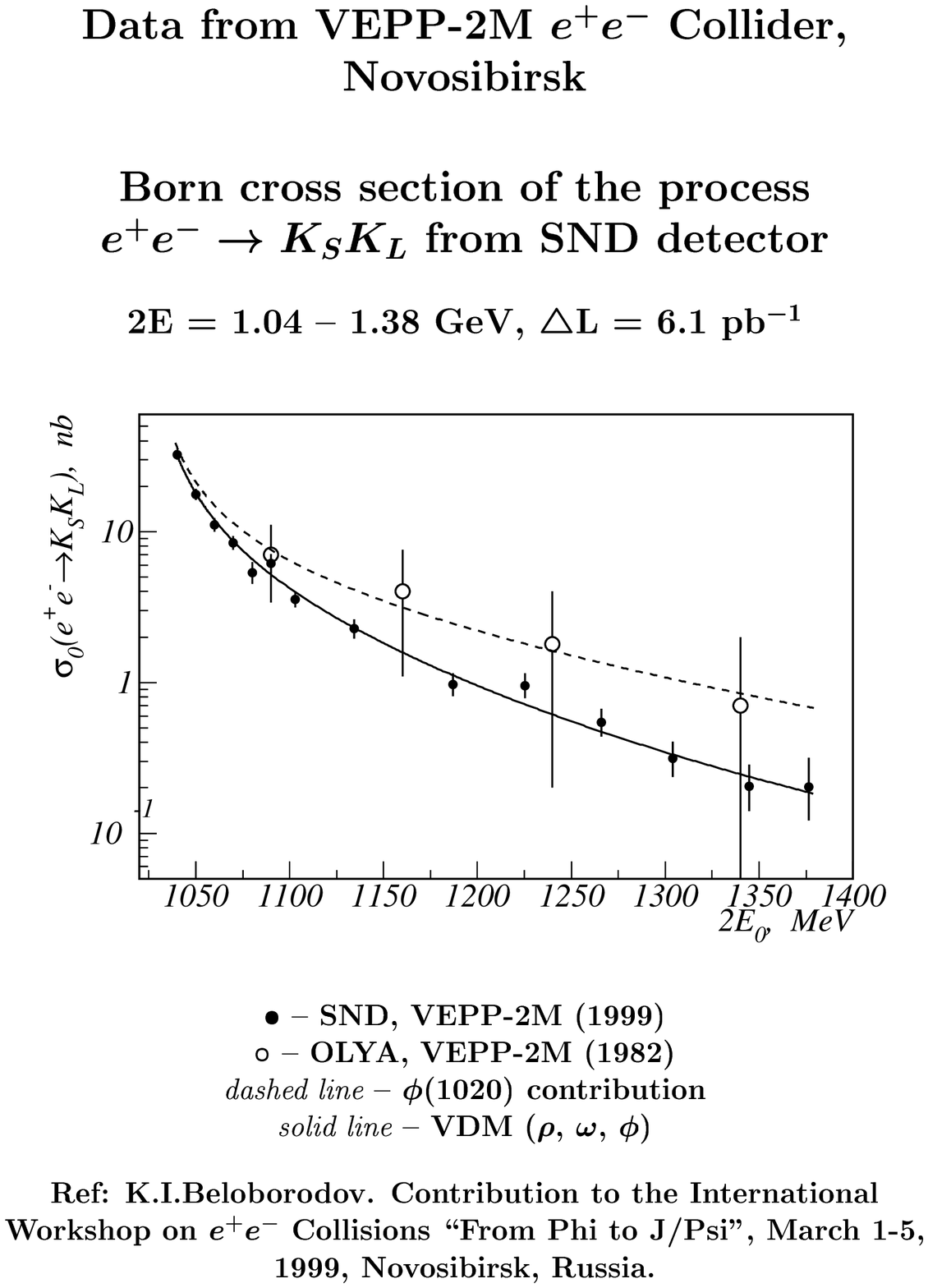,width=60mm,height=54mm,clip=true}}
    \caption[]{Energy dependence of the cross section of
               $e^+e^-\rightarrow K_S K_L$. Solid dots are from SND.}
    \label{fig:KsKL}
  \end{minipage}
\end{figure}

Invariant masses of $\pi$-meson pairs in the final $3\pi$ state were
measured to investigate the intermediate state in the $e^+e^-\rightarrow
\pi^+\pi^-\pi^0$ process. The intermediate states might be $\rho \pi$
and the much less probable $\omega\pi$ with decay $\omega\rightarrow
2\pi$.  Comparing the mass spectrum of $\pi^+\pi^-$ with that of $\pi^0
\pi^{\pm}$, one can observe the interference between the two
intermediate states.  The clear peak shown in $\pi^+\pi^-$ mass spectrum
of the experimental data proves the $\rho-\omega$ interference in $3\pi$
final state, and the phase measured to be zero agrees with the VDM
prediction.  Fig.~\ref{fig:pipipi0} plots the cross section for the
production of $\pi^+\pi^-\pi^0$.

%\begin{figure}[htb]
%\begin{center}
%\mbox{
%\psfig{figure=eps/pipipi0-ok.eps,width=60mm}}
%\mbox{
%\psfig{figure=eps/KsKL.eps,width=60mm,height=54mm,clip=true}}
%\parbox[l]{6.cm}{
%\caption{Energy dependence of the cross section of $e^+e^- \rightarrow
%\pi^+\pi^-\pi^0$. Solid line shows the VDM prediction.}
%\label{fig:pipipi0}}
%\hspace{2.5cm}
%\parbox{6.cm}{
%\caption{Energy dependence of the cross section of $e^+e^- \rightarrow
%K_S K_L$.}
%\label{fig:KsKL}}
%\end{center}
%\end{figure}

\noindent $\bullet$ Cross section measurement for $e^+e^-\rightarrow K_S K_L$

The cross section of $e^+e^-\rightarrow K_S K_L$ reaction was measured
in 1982 by OLYA in Novosibirsk\cite{Ivanov} and DM1 in Orsay\cite{Mane}
in the energy region $E_{cm}=1.06-1.40$ GeV and $E_{cm}=1.40-2.20$ GeV
respectively. It is desirable to re-measure this channel since the
accuracy reached by both experiments is poor.  So far, 1.8 pb$^{-1}$
data has been analyzed by SND, utilizing $K_S\rightarrow \pi^0 \pi^0$
from $e^+e^-\rightarrow K_S K_L$. $e^+e^-\rightarrow \omega
\pi^0\rightarrow \pi^0 \pi^0 \gamma$ is the main background source. In
addition, cosmic rays and beam associated background also
contribute. The cross section measured by SND is illustrated in
Fig.~\ref{fig:KsKL} with solid dots.

\noindent $\bullet$ $e^+e^-\rightarrow \omega\pi^+\pi^-$
($\omega \rightarrow \pi^+\pi^-\pi^0$), $\eta \pi^+ \pi^-$
($\eta \rightarrow \pi^+\pi^-\pi^0$, or $\gamma \gamma$) 

\begin{figure}[tbp]
  \begin{minipage}[t]{0.48\linewidth}
    \centerline{\psfig{file=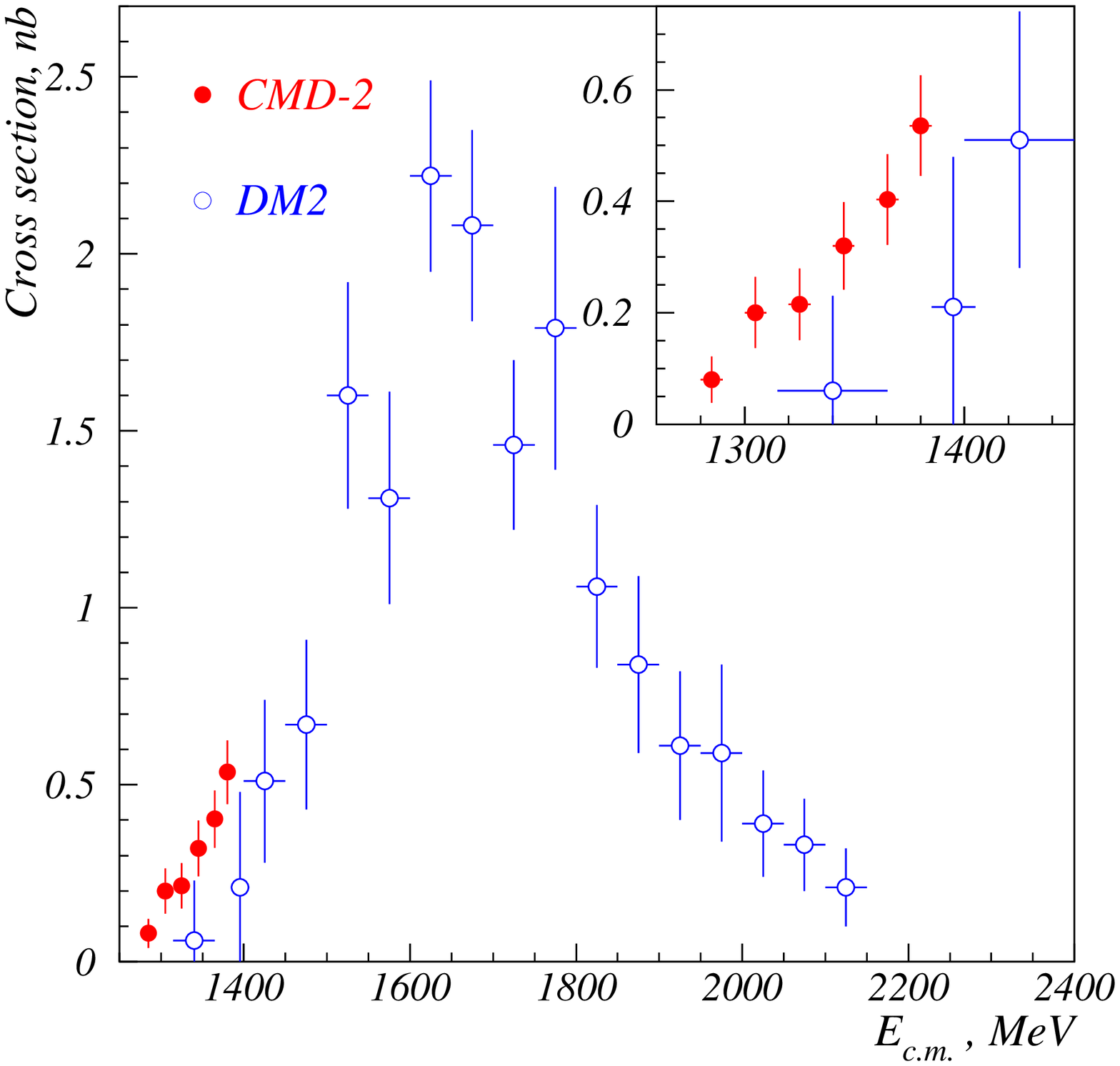,width=60mm}}
    \caption[]{Cross sections for the reactions $e^+e^-\rightarrow
               \omega \pi^+ \pi^-$($\omega\rightarrow \pi^+\pi^-\pi^0$).}
    \label{fig:omegapi}
  \end{minipage}
  \hfill
  \begin{minipage}[t]{0.48\linewidth}
    \centerline{\psfig{file=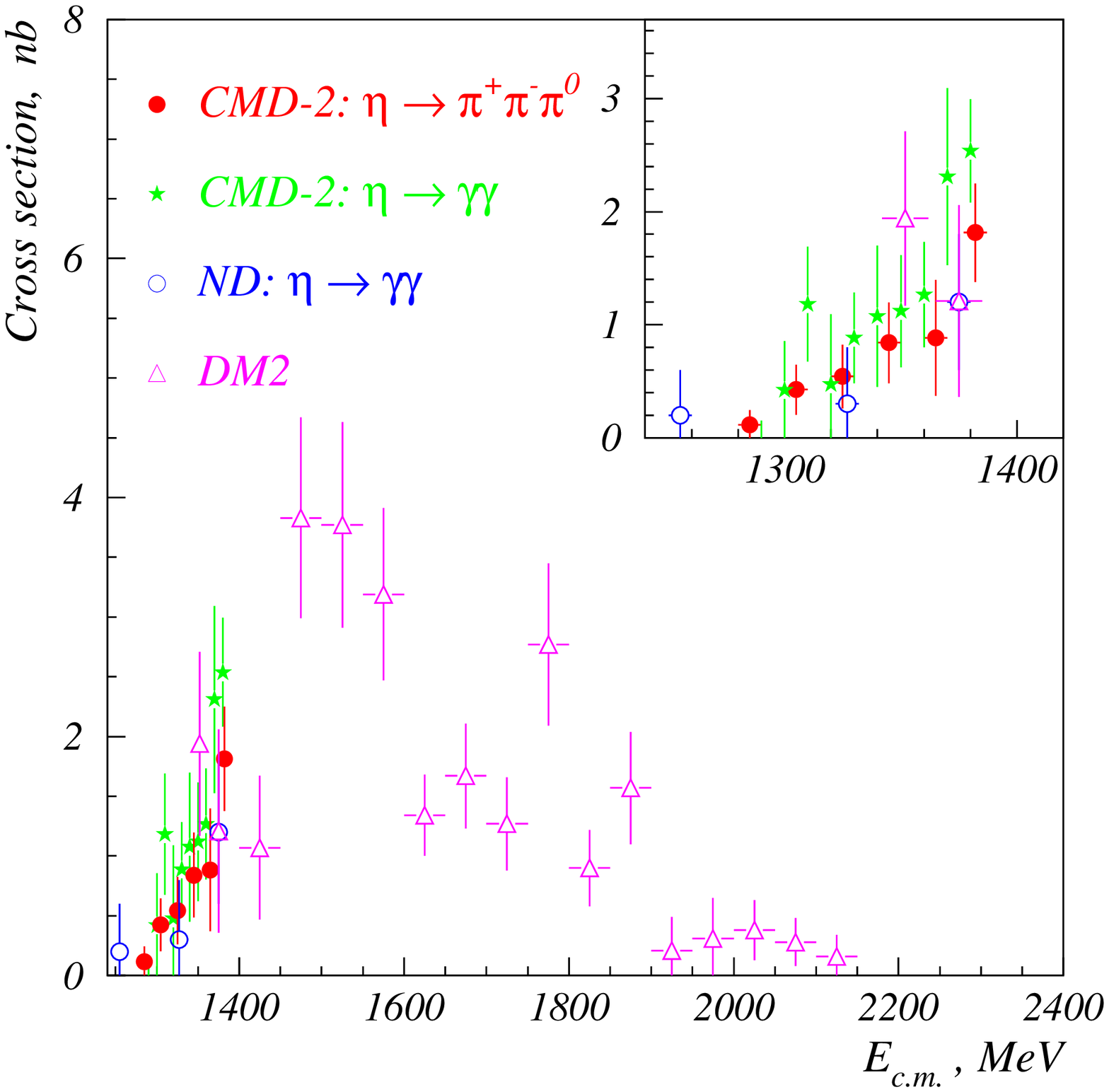,width=60mm,height=54mm,clip=true}}
    \caption[]{Cross section for $e^+e^-\rightarrow\eta\pi^+ \pi^-$
               ($\eta\rightarrow\pi^+\pi^-\pi^0$, or $\gamma \gamma$).}
    \label{fig:etapi}
  \end{minipage}
\end{figure}

Figs.~\ref{fig:omegapi} and ~\ref{fig:etapi} plot the cross section
measured by CMD-2 for $\omega \pi^+ \pi^-$ and $\eta \pi^+ \pi^-$,
together with the measurement by DM2.  The new measurement significantly
reduced the uncertainties though the systematic errors are still as high
as 15\%.

\noindent $\bullet$ $e^+e^-\rightarrow \pi^+ \pi^-$

The cross section of the process $e^+e^-\rightarrow \pi^+ \pi^-$ is
given by

\begin{equation} 
\sigma = \frac{\pi \alpha^2}{3s} \beta_{\pi}^3 |{F_{\pi}(s)}|^2, 
\end{equation}
where $F_{\pi}(s)$ and $\beta_{\pi}$ are the pion form factor at the cm
energy $\sqrt {s}$ and the velocity of the pion respectively.  A
precision measurement of the pion form factor is necessary to determine
the $R$-values via an exclusive method. The relative uncertainty
contribution to $a_{\mu}$ is dominated by the $e^+e^-\rightarrow \pi^+
\pi^-$ channel with $\sqrt{s} <2$~GeV~\cite{Eidelman,Brown}. The famous
E821 experiment at BNL\cite{LP99Lee} has measured $a_{\mu}$ to a
precision of $\sim5$~ppm and will further improve the accuracy to about
1 ppm. In order to compare a measurement with such a high accuracy with
theory, the uncertainty in $R$ should be below 0.5\% in this energy region.
Therefore a new measurement of the pion form factor with smaller
uncertainty is important for the interpretation of the E821 measurement.

The pion form factor was measured by the OLYA and CMD groups at VEPP-2M
about twenty years ago~\cite{Barkov}.  Twenty-four points from 360 to
820 MeV were studied by CMD with a systematic uncertainty of about 2\%.
The OLYA measurement scanned from 640 to 1400 MeV with small steps,
giving a systematic uncertainty from 4\% at the $\rho$-meson peak to
15\% at 1400 MeV.

The pion form factor is one of the major experiments planned at CMD-2.
A total of 128 energy points were scanned in the whole VEPP-2M
energy region (0.36-1.38 GeV) in six runs performed from 1994 to
1998~\cite{vepp2mscan}.  The discussion here is based on data taken from
the first 3 runs with 43 energy points ranging from 0.61-0.96 GeV.  The
small energy scan step, 0.01 GeV, in this energy region allows the
calculation of the hadronic contribution in a model-independent way.  In
order to investigate the $\omega$-meson parameters and the $\rho -
\omega$ interference, the energy steps were 2-6 MeV in the energy region
near the $\omega$-meson. The beam energy was measured with the resonance
depolarization technique for almost all the energy points, which
significantly reduced the systematic error arising from the energy
uncertainty. The charged trigger makes use of the information from the
drift chamber and the Z-chamber and requires at least one track. There
was an additional trigger criteria for the energy points between 0.81
and 0.96 GeV, which asks for the total energy deposited in the
calorimeter to be greater than 20-30 MeV. The neutral trigger, served
for monitoring the trigger efficiency, is based on the information only
from the calorimeter.

The background is mainly from cosmic muons. Bhabha and dimuon production
are also background sources. The shape of the energy deposition was
carefully studied for the event separation and selection.  An event
vertex cut was applied to reject cosmic muons effectively.

To account for the fact that the radiative correction for
$e^+e^-\rightarrow \pi^+ \pi^-$ depends on the energy behavior of the
cross section of $e^+e^-\rightarrow \pi^+ \pi^-$ itself,
the radiative correction factor was calculated iteratively. The existing
$|F_{\pi}(s)|^2$ data were used as the first iteration for the
calculation. The values of $|F_{\pi}(s)|^2$ were found to be stable 
after three iterations.

The corrections for the pion losses due to decays in flight and nuclear
interaction, as well as the background from $\omega\rightarrow 3 \pi$
were done using Monte Carlo simulations.

The total systematic uncertainty was estimated to be currently 1.5\% and
1.7\% for the energy region 0.78-0.784 GeV and 0.782-0.94 GeV
respectively, and 1.4\% for all the other points.
%Table ~\ref{tab:syserr} summarizes the main systematic 
%error sources.
%\begin{table}[h]
%\caption {Main sources of systematic errors}
%\vskip 0.3cm
%\begin{center}
%\begin{tabular}{|c|c|} \hline
%Source                         & Estimated value \\ \hline \hline 
%Events separation              & 0.6\% \\ \hline
%Energy calibration of collider & 0.1\%(0.5\% for 390,392; 1\% 391) \\ \hline
%Fiducial volume                & 0.5\%  \\ \hline
%Trigger efficiency             & 0.2\%(1\% for 420-435,470) \\ \hline
%Reconstruction efficiency      & 0.3\% \\ \hline
%Hadronic interactions of pions & 0.4\% \\ \hline
%Pions decays in flight         & 0.1\% \\ \hline
%Radiative correction           & 1\% \\ \hline \hline
%Total                          & 1.4\% \\ 
%                               &(1.5\% for 390,392) \\
%			       &(1.7\% for 391,420-435,470) \\ \hline 
%\end{tabular}
%\end{center}
%\label{tab:syserr}
%\end{table}

\begin{figure}[htb]
\centerline{
\psfig{figure=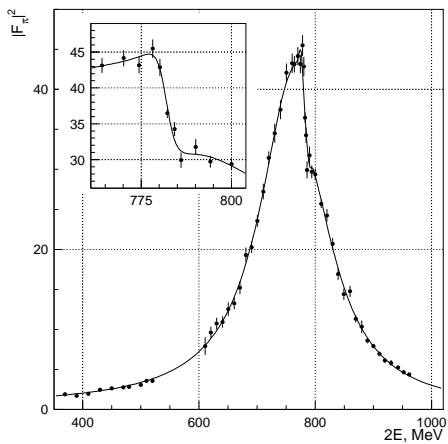,width=60mm,clip=true}}
\caption{pion form factor measured by CMD-2}
\label{fig:formfactor}
\end{figure}

Other than the leading contribution from $\rho(770)$ and $\omega(782)$,
the resonances $\rho(1450)$ and $\rho(1700)$ should be taken into
account to describe the data for the determination of the pion form
factor. In addition, the model based on the Hidden Local Symmetry (HLS),
which predicts a point-like coupling $\gamma \pi^+ \pi^-$, can well
describe the experimental data below 1 GeV. Both the Gounaris-Sakurai
(GS) parameterization and the HLS parameterization
approaches~\cite{Barkov,Barate} were used to fit the form factor.  Only
the higher resonance $\rho(1450)$ was taken into account in fitting the
pion form factor in the relatively narrow energy region 0.61-0.96 GeV.
Fig.~\ref{fig:formfactor} shows the fit of the pion form factor with
CMD-2 94, 95 data according to GS and HLS models. Both theoretical
curves are indistinguishable.

The pion form factor fit of CMD-2 94, 95 data is summarized by
Table~\ref{tab:formfit}. PDG data is also shown for comparison.

\begin{table}[h]
\caption {The results of fit of the CMD-2(94,95) pion form factor data
by GS and HLS models}
\vskip 0.3cm
\begin{center}
\begin {tabular}{|c|c|c|} \hline
               & GS model  & HLS model \\ \hline
$M_{\rho}(MeV)$   &775.28$\pm$0.61$\pm$0.20 & 774.57$\pm$0.60$\pm$0.20 \\ \hline
$\Gamma_{\rho}(MeV)$ &147.70$\pm$1.29$\pm$0.40 & 147.65$\pm$1.38$\pm$0.20 \\ \hline
$Br(\omega\rightarrow \pi^+\pi^-)$,\% &1.31$\pm$0.23$\pm$0.02 &
1.32$\pm$0.23$\pm$0.02 \\ \hline
$\beta$(GS)    &-0.0849 $\pm$ 0.0053 $\pm$0.0050 & - \\ \hline
$\alpha$(HLS)  & - & 2.381$\pm$0.016$\pm$0.016 \\ \hline
$\chi^2/n$     & 0.77   & 0.78 \\ \hline 
\end{tabular}
\end{center}
\label{tab:formfit}
\end{table}

With the remaining data collected, CMD-2 hopes to reduce the systematic
error presented here by a factor of two. To achieve this goal, a new
approach for the calculation of the radiative correction must be
developed.

%The new cross section measurements from Novosibirsk significantly
%improve the accuracy and are the precision measurements. It would be
%more contributing if the energy region could be extended to 2 GeV,
%which link up to the lowest energy of BEPC.

\subsection {$R$ scan with BESII at BEPC in Beijing}

With the upgraded machine and detector~\cite{BES1,BES2,rscan1}, the BES
collaboration performed two scans to measure $R$ in the energy region of
2-5 GeV in 1998 and 1999. The first run scanned 6 energy points covering
the energy from 2.6 to 5 GeV in the continuum. Separated beam running
at each energy point was carried out in order to subtract the beam
associated background from the data~\cite{rscan1}.

%The main goal of the first $R$ scan was to understand the 
%trigger and hadronic event acceptances, as well as the hadronic events 
%selection and the background subtraction, which are central to a total cross 
%section measurement. 

The second run scanned about 85 energy points in the energy region of
2-4.8 GeV\cite{rscan2}.  To subtract beam associated background,
separated beam running was done at 26 energy points and single
beam running for both $e^-$ and $e^+$ was done at 7 energy
points distributed over the whole scanned energy region. Special runs
were taken at the $J/\psi$ to determine the trigger efficiency. The 
$J/\psi$ and $\psi(2S)$ resonances were scanned at the beginning and 
at the end of the $R$ scan for the energy calibration.

The $R$ values from the BESII scan data are measured by observing the
final hadronic events inclusively, i.e. the value of $R$ is determined
from the number of observed hadronic events ($N^{obs}_{had}$) by the
relation

\begin{equation}
R=\frac{ N^{obs}_{had} - N_{bg} - \sum_{l}N_{ll} - N_{\gamma\gamma} }
{ \sigma^0_{\mu\mu} \cdot L \cdot \epsilon_{had} \cdot \epsilon_{trg}
\cdot (1+\delta)}, 
\end{equation}
where $N_{bg}$ is the number of beam associated background events;
$\sum_{l}N_{ll},~(l=e,\mu,\tau)$ and $N_{\gamma\gamma}$ are the numbers
of misidentified lepton-pairs from one-photon and two-photon processes
events; $L$ is the integrated luminosity; $\delta$ is the radiative
correction; $\epsilon_{had}$ is the detection efficiency for hadronic
events and $\epsilon_{trg}$ represents the trigger efficiency.

The trigger efficiencies are measured by comparing the responses to
different trigger requirements in special runs taken at the $J/\psi$
resonance. From the trigger measurements, the efficiencies for Bhabha,
dimuon and hadronic events are determined to be 99.96\%, 99.33\% and
99.76\%, respectively. As a cross check, the trigger information from
the 2.6 and 3.55 GeV data samples is used to provide an independent
measurement of the trigger efficiencies. This measurement is consistent
with the efficiencies determined from the $J/\psi$ data. The errors in
the trigger efficiencies for Bhabha and hadronic events are less than
0.5\%.

%\subsection{Hadronic event selection and background subtraction}
The task of the hadronic event selection is to identify one photon
multi-hadron production from all other possible contamination
mechanisms.  The event selection makes full use of all the information
from each sub-detector of BESII, namely, the vertex position, the
measured charged-particle momentum and the energy loss due to
ionization, the related time of flight, the associated pulse height 
and pulse 
shape of the electromagnetic calorimeter, and the hits in $\mu$ counter.

%Two categories of events will be selected, i.e. the lepton pair production
%$e^+e^- \rightarrow e^+e^-$, $\mu^+\mu^-$, $\tau^+\tau^-$ and the production 
%of three or more hadronic particles $e^+e^- \rightarrow hadrons$. 
%Events with three or more 
%prongs forming a vertex should be selected as hadronic events. Two prong
%events with total charge zero may be treated as hadronic events if the track 
%momentum and azimuthal angle are clearly out of the categories of the Bhabha 
%and dimuon events as described below.

The backgrounds involved in the measurement are mainly from cosmic rays,
lepton pair production ($e^+e^-$, $e^+e^-$, $\mu^+\mu^-$,
$\tau^+\tau^-$), two-photon processes, and beam associated processes.
The cosmic rays and part of the lepton pair production events are
directly removed by the event selection. The remaining background from
lepton pair production and two-photon processes is then subtracted out
statistically according to a Monte Carlo simulation.

%The lepton pair productions are QED processes and can be calculated to the
%precision of $O(\alpha^3)$ for $e$ and $\mu$ pair production.
%The $\tau^+\tau^-$ pair events are difficult to be separated from hadronic 
%events sample because of its hadronic decay and short life time, and should
%be therefore treated by a background subtraction with the help of Monte Carlo 
%simulation, in which all the possible decay modes and the corresponding
%branching ratio should be taken into account according to the PDG. The 
%fraction of events surviving the hadronic events criteria is energy 
%dependent. 
%and is high. It was about ?% and the overall magnitude of the 
%(-subtraction was about ?%, due to which the total contribution to the error 
%in R was estimated to be (?% [PRL R98].    
%Two-photon processes is a higher-order electromagnetic processes where both
%electron and positron emit a quasi real photon, the two photons giving rise
%to an object with mass m usually appreciably smaller than $\sqrt{s}$ and
%which decays through $e^+e^-\rightarrow l^+l^- \gamma$,
%$e^+e^-\rightarrow l^+l^-e^+e^-$, $e^+e^- \rightarrow l^+l^- \mu^+\mu^-$ and
%$e^+e^- \rightarrow e^+e^- hadrons(l=e, \mu)$. The cross section of
%the other two photon processes are small as compared to
%$e^+e^- \rightarrow l^+l^-e^+e^-$~\cite{twogamma,MarkI}. The typical features of
%the produced final particles via two-photon processes are that they are 
%pretty much favoring the direction along the beam pipe, and poses a very small
%momentum.

The beam associated background sources are complicated. They may mainly
come from beam-gas, beam-wall interaction, synchrotron radiation, and
lost beam particles.  The salient features of the beam associated
background are that their tracks are very much along the beam pipe
direction, the energy deposited in BSC is small, and most of the tracks
are protons.

%The cosmic rays and part of the lepton pair production events are 
%directly removed according to the vertex position, the time of flight and 
%the collinear angle of the selected tracks, as well as the associated hits
%on muon counter. The remaining background from the lepton pair production 
%and two-photon processes is then subtracted out statistically according to 
%Monte Carlo simulation. 

Separated-beam runs were performed for the subtraction of
beam associated background. Most of the beam associated background
events are rejected by vertex and energy cuts. Applying the same
hadronic events selection criteria to the separated-beam data, one can
obtain the number of separated-beam events $N_{sep}$ surviving these
criteria. The number of beam associated events $N_{bg}$ in the
corresponding hadronic events sample is given by $N_{bg} = f \cdot
N_{sep}$, where $f$ is the ratio of the product of the pressure at the
collision region times the integrated beam currents for colliding beam
runs and that for the separated beam runs.
%\begin{equation}
%N_{bg} = f \cdot  N_{sep},
%\end{equation}
%where f , the constant of proportionality, can be determined by the ratio of 
%the pressure at collision region times beam current integrated over time
%measured for both colliding- and separated-beam running, i.e.,
%\begin{equation}
%f = \frac{\int\limits_{run} \ P_{run} \cdot I_{run}\,dt}
%         {\int\limits_{sep} \ P_{sep} \cdot I_{sep}\,dt} ,
%\end{equation}
To subtract beam associated background in this way, the variation of the 
pressure in the collision region and the beam current must be recorded for 
both colliding and separated-beam runs at each energy to be measured.

JETSET7.4 is used as the hadronic event generator to determine the
detection efficiency for hadronic events.  Parameters in the generator
are tuned using a $4 \times 10^4$ hadronic event sample collected near
3.55 GeV for the tau mass measurement done by the BES
collaboration~\cite{BEStau}. The parameters of the generator are
adjusted to reproduce distributions of kinematic variables such as
multiplicity, sphericity, transverse momentum, etc.
%Fig.~\ref{fig:evtshape} shows these  distributions
%for the real and simulated event samples.
%\begin{figure}[htb]
%\centerline{
%\psfig{figure=eps/shape.eps,width=80mm}}
%\psfig{figure=eps/bese.eps,width=100mm}}
%\caption{Comparison of hadronic event shapes between data (shaded
%region) and Monte Carlo (histogram).  (a) Multiplicity; (b) Sphericity;
%(c) Rapidity; (d) Transverse momentum.}
%\label{fig:evtshape}
%\end{figure} 

The parameters have also been obtained using the 2.6 GeV data ($\approx
5 \times 10^3$ events).  The difference between the two parameter sets
and between the data and the Monte Carlo data based on these parameter
sets is used to determine a systematic error of 1.9-3.2\% in the
hadronic efficiency.

The Monte Carlo simulation packet JETSET was not designed to fully
describe few body states produced by $e^+e^-$ annihilation in the few
GeV energy region, though the event shapes are consistent with that from
the Monte Carlo simulation with tuned parameters at 3.5 GeV.  A great
effort has been made by the Lund group and BES collaboration to
develop the formalism using the basic Lund Model area law directly for
the Monte Carlo simulation, which is expected to describe the data
better~\cite{bo}.

%The radiative corrections for the initial state should be taken into account
%to obtain the true cross section from the experimentally measured one. To
%obtain the leading order $O(\alpha^2)$ cross section $\sigma^{0}_{had}$
%from the observed hadronic cross section $\sigma^{obs}_{had}$, one must
%remove the higher-order terms of the electromagnetic coupling constant
%$\alpha$, which is generally referred to as radiative corrections and is
%denoted by $\delta$  in our case. The observed cross section 
%$\sigma^{obs}_{had}$ is then related to $\sigma^{0}_{had}$ by 
%\begin{equation}
%\sigma^{obs}_{had}=\sigma^{0}_{had} \cdot \epsilon_{had} \cdot (1+\delta)
%\end{equation}

Radiative corrections determined using four different
schemes~\cite{radcorr1,radcorr2,radcorr3,radcorr4} agreed with each
other to within 1\% below charm threshold.  Above charm threshold, where
resonances are important, the agreement is within $1\sim3\%$.  The major
uncertainties common to all models are due to errors in previously
measured $R$-values and in the choice of values for the resonance
parameters. For the measurements reported here, we use the formalism of
ref.~\cite{radcorr3} and include the differences with the other schemes
in the systematic error of 2.2-4.1\%.

\begin{figure}[htbp]
\centerline{
\psfig{figure=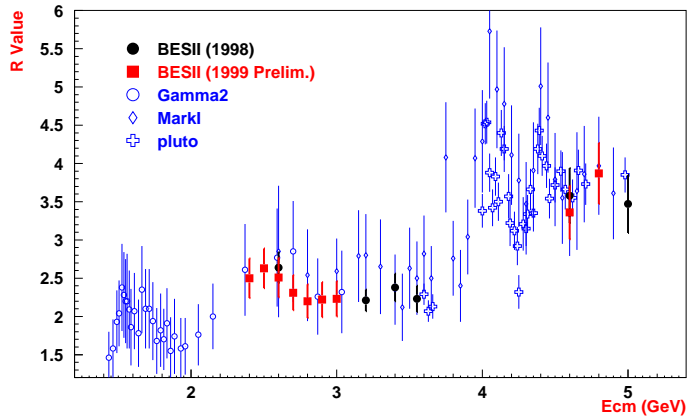,width=90mm,angle=270,clip=true}}
\caption{Plot of $R$ values vs $E_{cm}$}
\label{fig:rvalue}
\end{figure}

The $R$ values obtained at the six energy points scanned in 1998 are
shown in Table 3 
%~\ref{tab:r1998} 
and graphically displayed in
Fig.~\ref{fig:rvalue} with solid dots.
%Table~\ref{tab:rsyst} illustrates the systematic errors from 
%different error sources. 
The largest systematic error is due to the hadronic event selection and
is determined to be 3.8-6.0\% by varying the selection criteria. The
systematic errors on the measurements below 4.0 GeV are similar and are
a measure of the amount of error common to all points.  The BES
collaboration has also performed the analysis including only events with
greater than two charged tracks; although the statistics are smaller,
the results obtained agree well with the results shown here.

%\begin{table}[h]
%\begin{center}
%\caption{Summary of $R$ data and values.}
%\vskip 0.3cm
%\begin{tabular}{|c|c|c|c|c|c|c|c|c|} \hline
%$E_{cm}$ & $N_{had}^{obs}$ & $N_{bg}$ & $L$  &
%$\epsilon_{had}$ & $(1+\delta)$ & $R$  & Stat.  & Sys.  \\
%(GeV)  & & & (nb$^{-1})$ & (\%) & &  & error & error \\ \hline
%2.60 & 5617 & 127 & 292.9 & 54.82 & 1.009 & 2.61 & 0.05 & 0.17 \\
%3.20 & 2051 & 100 & 109.3 & 64.30 & 1.447 & 2.26 & 0.07 & 0.13 \\
%3.40 & 2149 & 178 & 135.3 & 69.61 & 1.173 & 2.37 & 0.07 & 0.16 \\
%3.55 & 2672 & 216 & 200.2 & 68.40 & 1.125 & 2.30 & 0.06 & 0.16 \\
%4.60 & 1497 & 282 &  87.7 & 82.27 & 1.079 & 3.56 & 0.20 & 0.29 \\
%5.00 & 1648 & 463 & 102.3 & 84.53 & 1.068 & 3.45 & 0.32 & 0.29 \\ \hline
%\end{tabular}
%\end{center}
%\label{tab:rvalue}
%\end{table}

\begin{table}[htb]
\begin{center}
%\caption{$R$ values and the contributions to systematic errors: hadronic 
%selection, $f$ factor, luminosity determination, $\tau$-pair background,
%background from Bhabha events, hadronic efficiency determination,
%trigger efficiency, and radiative corrections.  $\Delta R$ is the absolute
%error of $R$, which add the systematic and statistic errors in quadrature. 
%All other errors are in percentages (\%). }
%\vskip 0.3cm
%\begin{tabular}{|c|c|c|c|c|c|c|c|c|c|c|} \hline
%$E_{cm}$ & $R$  & $\Delta R$ & Had. & $f$ & $L$ & $\tau$-pair & Bhabhas
%& Had. & Trig. & Rad. \\
%(GeV)  & & & sel. & factor &  &  &  & eff. & & corr. \\ \hline
%2.60 & 2.64 & 0.18 & 5.1 & 0.06 & 2.12 & 0.00 & 0.04 & 2.60 & 0.50 & 2.6 \\
%3.20 & 2.21 & 0.15 & 3.8 & 0.15 & 2.83 & 0.00 & 0.04 & 1.90 & 0.50 & 2.2 \\
%3.40 & 2.38 & 0.17 & 4.6 & 0.27 & 2.83 & 0.00 & 0.04 & 2.90 & 0.50 & 3.0 \\
%3.55 & 2.23 & 0.17 & 5.5 & 0.27 & 2.32 & 0.00 & 0.04 & 2.30 & 0.50 & 2.4 \\
%4.60 & 3.58 & 0.35 & 5.7 & 0.75 & 2.16 & 0.32 & 0.00 & 3.60 & 0.50 & 4.1 \\
%5.00 & 3.47 & 0.43 & 6.0 & 1.26 & 2.81 & 0.32 & 0.00 & 3.20 & 0.50 & 3.8 \\ \hline
\caption{$R$-values measured with BESII from 1998 run.}
\vskip 0.3cm
\begin{tabular}{ccccccc}
\hline
$E_{cm}(GeV)$ & 2.6   & 3.2   & 3.4  & 3.55 & 4.6  & 5.00 \\\hline
$R$           & 2.64  & 2.21  & 2.38 & 2.23 & 3.58 & 3.47 \\
Stat. error   & 0.05  & 0.07  & 0.07 & 0.06 & 0.20 & 0.32 \\
Sys. error    & 0.19  & 0.13  & 0.16 & 0.16 & 0.29 & 0.29 \\\hline
\end{tabular}
\end{center}
\label{tab:r1998}
\end{table}

The $R$ values for $E_{cm}$ below 4 GeV are in good agreement with
results from $\gamma \gamma 2$ \cite{gamma2} and Pluto \cite{pluto} but
are below those from Mark I \cite{MarkI}. Above 4 GeV, our values are
consistent with previous measurements.

\begin{figure}[htb]
\centerline{
\psfig{figure=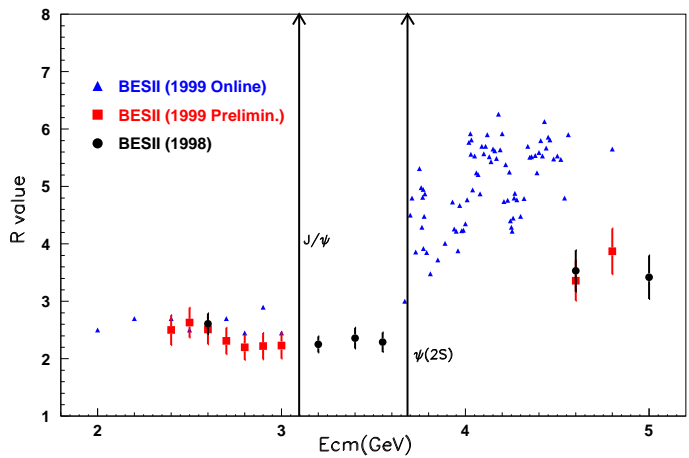,width=90mm,angle=270,clip=true}}
\caption{Energy points scanned with BESII. $R$ values shown with
triangles are from the online data estimation, of which the detection
efficiency, the raditive correction and the background have not yet been
taken into account. The solid sqaures represent the preliminary $R$-values
obtained from BESII data, of which the total errors are all
conservatively assigned to be 10\%. The final $R$-values of the 6 points
listed in table 3 are also plotted with solid dots for comparison.}
\label{fig:r99}
\end{figure}

Preliminary $R$-values at 2.4, 2.5, 2.6, 2.7, 2.8, 2.9, 3.0, 4.6 and 4.8
GeV are plotted with solid squares in Fig.~\ref{fig:r99}. 
The preliminary errors, which add the
statistical and systematical errors in quadrature, are all
conservatively assigned to be 10\%. However, it
is believed that these errors can be decreased to be comparable to 
the error bars of the solid dots, i.e. $\sim7\%$ for the
energy points below 3.6 GeV and $\sim10\%$ for energies above 4.5
GeV. The first scan repeated 3.4 GeV and the second scan repeated 2.6
and 4.6 GeV data points measured in the first scan. In all case,
$R$-values obtained are consistent with each other at the same energy.

%The $R$ scan done with BESII were well planned, and the performance was stable
%and the data quality is good. A great effort was invested to understand and 
%improve the detection efficiency, to determine the trigger efficiency and to 
%correctly subtract the beam associated background from the data with the help of 
%the separated/single beam operation data. The uncertainty in R measured by BES in the 
%energy region below 3.6 GeV will be decreased a factor of two, reaching 7\% accuracy. 
%The accuracy is not expected to be better than 10\% for the energy above 4 GeV.

%The new R ratio in $e^+e^-$ presented from Novosibirsk and Beijing has a great impact
%on the improvement of $\alpha(M_Z^2)$. Using these new results, though some of them
%are still preliminary,  A.D. Martin et. al. \cite{Martin} re-evaluate $\alpha(M_Z^2)$ 
%and find $\alpha(M_Z^2)^{-1}=128.973 \pm 0.035$ or $128.934 \pm 0.040$,
%according to whether inclusive or exclusive cross section are used. With the
%final results from Beijing inclusive measurement in the whole 2-5 GeV region and
%the more precise results from Novosibirsk, $\alpha(M_Z^2)$ and $(g-2)_{\mu}$ will 
%be more precisely determined from the experimental data.

\section{Prospects and Concluding Remarks}

SND and CMD-2 at VEPP-2M have significantly improved the measurements of
the hadron production cross section via $e^+e^-$ collisions for some of
the important exclusive channels in the energy region of 0.36-1.38 GeV.
Further improvement with the analysis of the existing data is
forthcoming.  A major advance would be possible if the energy
region could be extended to 2 GeV, which would link up to the lowest
energy of BEPC.

CMD-2 and SND at VEPP-2M are planning to scan from threshold to 1.4 GeV
in 1999-2000. A $R$ scan between 2 to 10 GeV with KEDR at VEPP-4 is
proposed. A scan covering such a wide energy region with the same
machine and detector would be very important if the measurement could be
performed with a $\sim1\%$ precision.

%In addition, there are proposals to build $\phi$
%and $\tau$-c factory. Both would be nice tools to measure R precisely.

The $R$ scan performed with BESII at BEPC in Beijing can significantly
reduce the uncertainties in $R$ in the energy region 2-5 GeV. The
$R$-values from the first run data have already reduced the
uncertainties in $R$ from 15-20\% to 7\%. BESII at BEPC in Beijing is
analyzing the second run $R$ scan data. The preliminary $R$ values in
the whole energy region of 2-5 GeV are expected to be shown in the
Spring of 2000, and the final results will be presented in the summer of
2000.

%The Chinese government has significantly increase the budget for the operation 
%of BEPC/BESII. In addition, 8 million US dollars has been approved to 
%BEPC/BESII for their major upgrade and for the R\&D for the Beijing 
%$\tau$-charm factory(BTCF). If the BTCF dream become true, the R values 
%in the energy region 2-5 GeV, particularly in 2-3.7 GeV could be measured 
%to an accuracy of 1-3\%, which would be extremely important for the 
%interpretation of $(g-2)_{\mu}$ experiment carrying out by E821 at BNL 
%and for  the precision determination of $\alpha(M_Z^2)$.
%\section{Concluding Remarks}

The new $R$ ratio results in $e^+e^-$ annihilations presented from
Novosibirsk and Beijing have had a great impact on the value of
$\alpha(M_Z^2)$. Using these new (albeit, preliminary) results,
A.D. Martin et. al. \cite{Martin} re-evaluate $\alpha(M_Z^2)$ and find
$\alpha(M_Z^2)^{-1}=128.973 \pm 0.035$ or $128.934 \pm 0.040$, according
to whether inclusive or exclusive cross sections are used.  The
uncertainties here are already decreased more than half if we compare 
with the previous value of $\alpha(M_Z^2)^{-1}=128.89 \pm 0.09$ evalued 
by using the old experimental $R$-values~\cite{Blondel}.  
With the final results from the Beijing inclusive measurement in the 
whole 2-5 GeV region and the more precise results from Novosibirsk,
$\alpha(M_Z^2)$ and $a_{\mu}$ will be even more precisely determined 
from the experimental data.

A dedicated energy scan, aimed at a 1\% precision direct measurement
below 1.4 GeV, planned by KLOE at DAPNE is not possible in the short 
term since the DAFNE machine is tuned for the $\phi$ resonance. 
However a
machine upgrade is foreseen which will hopefully permit such an energy
scan around 2004.  Another method to measure the hadronic cross section
is to measure events with Initial State Radiation (ISR).  In this case
one of the electrons or positrons of the beam radiates a hard photon and
the cm energy of the hadronic system in the final state, mostly pions
coming from the $\rho$-resonance, is lowered. KLOE has already started
the analysis of those events~\cite{RKLOE}.
%KLOE at DAPNE in Frascati can potentially improve the R-values to aprecision
%of 1\% in the energy region from the hadron production threshold to 1.4 
%GeV. An R scan
%extending the energy from the threshold to 2 GeV, which links up the scan
%energy to
%the lowest done with BESII at BEPC, is highly wished to improve the measurement.

Being one of the most fundamental parameters in particle physics, the
$R$-value plays an important role in the development of the theory of
particle physics and in testing the Standard Model.  Experimental
efforts to precisely measure $R$-values at low energies are crucial for
the future electroweak precision physics. The measurements are not only
important for the evaluation of $\alpha(M_Z^2)$ and for the
interpretation of $a_{\mu}$, but also necessary for the understanding of
the hadron production mechanism via $e^+e^-$ annihilation.
% Rest of this sentence is awkward. -Derrick
%using the data with enough statistics collecting in the continue region
%below 5 GeV.

A real breakthrough in electroweak theory physics with regard to the
$R$-values at low energy would be possible only by measuring
$\sigma(e^+e^-\rightarrow \mbox{hadrons})$ at $\sim1\%$ accuracy. Such a
level of precision requires significant improvement to both machine and
detector, and needs better theoretical calculation of the radiative
correction and the event generator for hadron production.

Once the $R$-values has been measured with a precision of ~1\% in the
energies covered by VEPP-2M and DAPNE, the central question will be
how to further decrease the uncertainties of $R$-values measured by
BESII in the energy region of 2-5 GeV, particularly from 2-3.7 GeV.
Such a measurement would be extremely important for the interpretation
of the $a_{\mu}$ experiment carrying out by E821 at BNL and for the
precision determination of $\alpha(M_Z^2)$. This measurement would then
be an important and attractive physics program for a $\tau$-c factory.
%or any advanced $\tau$-c facility with a luminosity of $10^{32}$
%cm$^{-2}$s$^{-1}$ at the $J/\psi$ resonance and a new generation
%detector as compared to BESII.

I would like to thank the many people who have at some point or another
helped realize this presentation. In particular, my very special thanks are
due Dr. Boris I. Khazin, Dr. Sergey Serednjakov for supplying me useful
informations concerning the results from CMD-2 and SND.  I'm also gratful 
to Dr. W. Dunwoodie, Dr. A. F. Haris and Dr. D. Kong who gave me helpful
comments and suggestions.


\begin{thebibliography}{99}
\bibitem{Rhighorder} S.G. Gorishny {\it et al.}, \Journal{\PLB}{ 259}{(1991)}
{80}
\bibitem{RtoZ} C. Caso et al., {\em Eur. Phys. Jour.} C{\bf~3}(1998){1}.
%               R. Marshall \Journal{\ZPC}{43}{ 595}. 
\bibitem{Rlow1} F. Ceradini {\it et al.}, \Journal{\PLB}{ 47}{(1973)}{80};\\
            B. Bartoli {\it et al.}, \Journal{\PRD}{ 6}{(1972)}{2374};\\
            M. Bernardini {\it et al.}, \Journal{\PLB}{ 51}{(1974)}{200}.
\bibitem{Rlow2} G. Cosme {\it et al.}, \Journal{\PLB}{ 40}{(1972)}{685}.
\bibitem{Rlow3} M. Kurdadze {\it et al.}, \Journal{\PLB}{ 42}{(1972)}{515}.
\bibitem{Rlow4} A. Litke {\it et al.}, \Journal{\PRL}{ 30}{(1973)}{1189}.
\bibitem{gamma2} C. Bacci {\it et al.}, ($\gamma \gamma2$ Collab.),
\Journal{\PLB}{ 86}{(1979)}{234}.
\bibitem{MarkI} J. L. Siegrist {\it et al.}, (Mark I Collab.), \Journal
{\PLB}{ 26}{(1982)}{969}.
%\bibitem{Crystalball} A. Osterheld {\it et al.}, SLAC-PUB-4160;
%C. Edwards \etal, SLAC-PUB-5160.
\bibitem{pluto} L. Criegee and G. Knies, (Pluto Collab.),
{\em Phys. Rep.} {\bf 83}(1982)151;\\
Ch. Berger {\it et al.}, \Journal {\PLB}{ 81}{(1979)}{410}.
\bibitem{Glashow} S. L. Glashow, \Journal{\NP}{ 22}{(1961)}{579}.
\bibitem{Weinberg} S. Weinberg, \Journal{\PRL}{ 19}{(1967)}{124}.
\bibitem{Jegerlehner} F. Jegerlehner, \Journal{\ZPC}{ 32}{(1986)}{195} and
hep-ph/9606484.
\bibitem{DASP} R. Brandelik {\it et al.}, \Journal{\PLB}{ 76}{(1978)}{361}.
\bibitem{MarkIres} J. L. Siegrist {\it et al.},
\Journal{\PRD}{ 26}{(1982){969}.
\bibitem{PLUTOres} J. Burmeister {\it et al.},
\Journal{\PLB}{ 66}{(1977)}{395}.
\bibitem{BP} H. Burkhardt and B. Pietrzyk, \Journal{\PLB}{ 356}{(1995)}{398}.
\bibitem{Bolek} B. Pietrzyk, Proc. of the 3rd Int. Symp. on Radiative
Corrections, Cracow, Poland, Aug. 1996, Acta. Pol. B{\bf 28}(1997)673.
\bibitem{Davier1} M. Davier and A. H\"{o}ker, \Journal{\PLB}{ 419}
{(1998)}{419}.
\bibitem{Eidelman} S. Eidelmann and F. Jegerleher,
\Journal{\ZPC}{ 67}{(1995)}{585}
%H. Burkhardt and B. Pietrzyk, \Journal{\PLB}{356}{(1995)}{ 398}
\bibitem{swartz} M. L. Swartz,  \Journal{\PRD}{ 53}{(1996)}{5268}.
\bibitem{Davier2} R. Alemany, M. Davier, and A. H\"{o}ker, {\em Eur. Phys. Jour.}
C{ \bf~2}(1998)123.
\bibitem{Blondel} A. Blondel, Proc. of the 28th Int. Conf. on High Energy
Physics, Warsaw, Poland, July 1996.
\bibitem{g21} J. Bailey {\it et al.}, \Journal{\PLB}{ 68}{(1977)}{361}.
\bibitem{g22} F.J.M. and E. Picasso, Ann. Rev. Nucl. Sci. 29(1979)243
\bibitem{g23} F.J.M. Farley, \Journal{\ZPC}{ 56}{(1992)}{S88}
%\bibitem{g2rad1} D.J. Broadhurst, {\it et al.}, \Journal{\PLB}{ 298}
%{(1993)}{ 445} 
%\bibitem{g2rad2} T. Kinoshita, \Journal{\PRD}{47}{(1993)}{ 5013}. 
%\bibitem{g2SM1} T. Kinoshita and W.J. Marciano, in "Quantum Electrodynamics",
%ed. T. Kinoshita, World Scientific, Singapore, 1990, pp. 419-478
%\bibitem{g2SM2} P. Mery, {\it et al.}, \Journal{\ZPC}{46}{(1990)}{ 229}.
%\bibitem{g2SM3} W. Bernreuther, {\it et al.}, \Journal{\ZPC}{56}{(1992)}{ S97}.
\bibitem{Lee} B. Robert Lee (BNL E821), {\it et al.},
\Journal{\ZPC}{ 56}{(1992)}{S101}. 
\bibitem{bo} Bo Andersson and H.M. Hu, "Few-Body States in Lund String
Fragmentation Model", submitted to Eur. Phys. Jour.
\bibitem{vepp2mscan} R. R. Akhmetshing {\it et al.}, hep-exp/9904027,
26 April 1999.
\bibitem{snddet} V.M. Aulchenko {\it et al.}, Proc. Workshop on Physics 
and Detectors for DA$\phi$NE, Frascati, Italy, April 9-12(1991)605;\\
M. N. Achasov {\it et al.}, hep-ex/9909015, submitted to Nucl. Instr. And Meth.
\bibitem{sndresult} M. N. Achasov {\it et al.}, hep-exp/9809013, 16 Sep
1998. 
\bibitem{hadronspe} PDG Review of Particle Physics. Part I and II;\\ 
                    Phys. Rev. D, Particles and Fields.V.54 (1996).
\bibitem{snd15} S.I. Serednyakov {\it et. al.}, Phys. Rep. 202(1991)99.
\bibitem{snd47} L.M. Kurdardze {\it et. al.}, JETP Lett. 47(1988)512.
\bibitem{snd48} L.M. Kurdardze {\it et. al.}, JETP Lett. 43(1986)643.
\bibitem{snd49} D. Bisello {\it et. al.}, preprint LAL 90-35(1990).
\bibitem{snd50} C. Bacci {\it et al.}, \Journal{\NPB}{ 184}{(1981)}{31}.
\bibitem{dolinsky} S.I. Dolingdky {\it et. al.}, Phys. Rep. 202(1991)99.
\bibitem{Ivanov} P.M. Ivanov {\it et. al.}, Pisma v JETP, 36(1982)91.
\bibitem{Mane} F. Mane {\it et al.}, \Journal{\PLB}{ 99}{(1981)}{261}.
\bibitem{cmd2det} R. R. Akhmetshing {\it et al.}, Preprint BINP 99-11, 
Novosibirsk, 1999.
\bibitem{Brown} D.H. Brown and W.A. Worstell, \Journal{\PRD}{ 54}
{(1996)}{3237}.
\bibitem{olya} R. Barate {\it et al.}, \Journal{\ZPC}{ 76}{(1997)}{15}.
\bibitem{Benayoun} M. Benayoun {\it et al.}, Eur. Phys. J. C2(1998)269.
\bibitem{cmd223} G.J. Gounaris and J.J. Sakurai,
\Journal{\PRL}{ 21}{(1968)}{244}.
\bibitem{cmd224} M. Bando et al., \Journal{\PRL}{ 54}{(1985)}{1215}.
\bibitem{LP99Lee} B. Robert Lee, '$g-2$' talk given at LP99, Stanford, July 
27-Augest 3, 1999.
\bibitem{Barkov} l.M. Barkov {\it et al.}, \Journal{\NPB}{ 256}{(1985)}{365}.
\bibitem{Barate} R. Barate {\it et al.}, {\em Eur. Phys. Jour.} C{\bf 76}
(1997)15.
\bibitem{BES1} J.Z. Bai {\it et al.}, (BES Collab.),
\Journal{\NIMA}{ 344}{(1994)}{319}.
\bibitem{BES2} `The Upgraded BES Detector (BESII)'' talk by J. Li at
Workshop on Relativistic Heavy Ion Collisions and Quark Matter
Physics, Wuhan, China, April 5-8, 1999.
\bibitem{rscan1} Z.G. Zhao, CERN COURIER Vol.38 Number 6 Sep. 1998;\\ 
'Measurement of R with BESII at BEPC', a talk presented by Z.G. Zhao at
ICHEP98, Vacouver, 1998.\\
`Measurement of $R$ Between 2-5 GeV'' a talk
presented by D. Kong, DPF99 at Los Angeles, California, Jan. 5-9, 1999;
hep-ph/9903521.
\bibitem{rscan2} Z.G. Zhao and F.A. Harris, CERN COURIER 
Vol.39 Number 7 Sep. 1999.
\bibitem{twogamma} J. Parisi {\it et al.}, \Journal{\PRD}{ 4}{(1971)}{2927};\\
                   R. Brandelik {\it et al.}, \Journal{\PLB}{76}{(1978)}{361}.
\bibitem{lund} T. Sjostrand, Computer Phys. Commun. 82(1994)74;\\
               B. Andersson {\it et al.}, \Journal{\ZPC}{ 20}{(1983)}{317};\\
               B. Andersson, 'The Lund Model', Cambridge University Press,1988.
\bibitem{BEStau} J.Z. Bai {\it et al.}, \Journal{\PRD}{ 53}{(1996)}{20};\\
                 X.R. Qi {\it et al.}, High Energy and Nucl. Phys.
23(1999)1-9, in Chinese.
\bibitem{radcorr1} F.A. Berends and R. Kleiss,
\Journal{\NPB}{ 178}{(1981)}{141}.
\bibitem{radcorr2} E. A. Kuraev {\it et al.}, {\em Sov. J. Nucl. Phys.} 
{ \bf 41}(1985)3.
\bibitem{radcorr3} G. Bonneau and F. Martin, \Journal{\NPB}{ 27}{(1971)}
{387}.
\bibitem{radcorr4} C. Edwards {\it et al.}, SLAC-PUB-5160, 1990. 
\bibitem{Martin} A.D. Martin {\it et al.}, hep-ph/9912252
\bibitem{RKLOE} G. Cataldi {\it et al.}, KLOE MEMO \#195 Augest 13, 1999. edc
99-24. 
}
\end{thebibliography}
\end{document}